*Original Article*

# Trust Aware Privacy Preserving Routing Protocol for Wireless Adhoc Network

B. Murugeshwari[1], D. Saral Jeeva Jothi[2], B. Hemalatha[3], S. Neelavathy Pari[4]

[1,2,3]*Velammal Engineering College, Chennai, India*
[4]*Department of Computer Technology, MIT Campus, Anna University, Chennai, India*

[1]*Corresponding Author : niyansree@gmail.com*

*Abstract - Wireless Ad-Hoc Networks are especially helpful and quite well for essential circumstances such as defense, public safety, and disaster recovery. MANETs require communication privacy and security, notably in core routing protocols, when functioning in hostile or suspicious environments. The Trust Aware Privacy-Preserving Protocol (TAP3) is a mechanism for supporting the origin in proactively selecting a trust-able target and doing privacy-preserving route verification. We suggest TAP3 using the fellow recommendation model for MANETs in this work. Nodes use their features to discover their fellow node and use the trust to create strong connections with the random node via a multi-hop trusting chain by identifying the secure location. The verification duties are then spread among the nodes and validate the log updates without exposing the nodes' details. Unlike previous models that uncover node vulnerabilities or misconduct after an attack, TAP3 may guarantee the origin node to prevent data from being transferred through malicious nodes from the beginning and do verification without needing a third party. Our results show that this approach can locate problematic nodes with minimal overhead than the conventional routing protocol.*

## 1. Introduction

To improve communication mobility, fourth-generation (4G) wireless communication combines mobile ad hoc networks (MANET) with other connections such as cell technology, wireless personal area networks, and third-generation (3G) networks. The primary purpose of the 4G network is to enable mobile nodes to migrate around the world without even being constrained by enabling infrastructure [1-3]. The 4G systems provide one of the newer wireless networks known as MANETs. MANET is a mobile node network that uses multi-hop wireless transmitting and can operate without centralized infrastructure. Because wireless ad hoc lacks a stable infrastructure, nodes increasingly depend on fellow nodes for interaction [4]. The nodes can configure individually and construct an ad hoc architecture on the move.

Moreover, many MANET implementation situations include functioning in dangerous conditions, implying that assaults are either anticipated or possible at the very minimum [5]. Whereas most previous work in protected MANET route discovery concentrated on security problems, less attention to privacy. Note that privacy doesn't mean confidentiality of interaction (i.e., data) between many MANET endpoints; that's also a fundamental aspect of protected MANET operation. Cryptography quickly acquires suitable access control remedies to establish or maintain the network.

Integrating MANETs to the unsecured network for web access, on the other hand, poses significant risks and obstacles [10,11]. Ad hoc networking technologies typically have different compatibility than traditional internet routing algorithms. In ad hoc networks, routing protocols help with route training and management, whereas the web handles these activities by specialized routers executing routing algorithms. The communication between web nodes and mobile ad hoc networks is managed by customized mobile gates (MG) positioned at the MANET's border and linked to the communication infrastructure and the MANET. The MG must execute the infrastructural network's routing mechanism and the MANET's ad hoc routing algorithm to offer an interconnection between the two or more networks.

Nevertheless, these previous mechanisms cannot be employed in the decentralized and simultaneous secure route-finding process. Initially, the dynamic routing protocol's connection overhead typically increases as the network size grows. In contrast, reactive routing approaches without a safe technique will result in some transmission errors when malicious nodes are present. Furthermore, many routes, including AODV, DSR, and Multicast, use source-based navigation. After sending the RREQ, the origin does not influence data transmission until a path to the target. As a result, various assaults, such as the black hole attack [6] and the wormhole attack [7], might occur during the route discovery process. Most significantly, they fail miserably to validate the behavior of nodes anywhere along the system's chosen path. The source has no way of knowing whether the subsequent routes have securely delivered the necessary communications and performed as intended without the participation of vicious attackers. The major problem for multi-path algorithms is choosing the path that decreases node failure probability while extending the network's lifespan [8]. During the routing discovery step, selecting a friendly approach and performing the verification are required.

This study offers TAP3 confirmation in MANETs, a dynamic direction-finding exploration, and an automatic authentication process to address the issues above. For starters, using active learning, TAP3 can assist the source in discovering the actual destination. The method then verifies independently throughout the route to determine whether the intermediary routers are fraudulent. Our dynamic training describes a systematic selection theory [9] determining the multivariate vector length between the targeted host's present and typical states. Apart from that can be performed in a disseminated manner without assembling the entire node's data. Throughout authentication, the decentralized nodes work together to extract evidence from the route log using preset reasoning principles. Nodes may identify suspicious nodes upon that path from source to destination and their actual location using the obtained proof. Finally, TAP3 does not jeopardize the objective of saving every node's privacy [21]. Joining log tables from multiple routers is unnecessary to browse through several network log entries. Instead of relying on log proof acquired during the execution stage, TAP3 uses a combination of analysis and verification to fight attackers.

## 2. Related Work

Sun et al. [12] presented Daemon among the first solutions for integrating the ad hoc and Mobile network routing mechanisms. However, this method makes no mention of privacy or security measures. Using various routing algorithms for path-optimized connectivity across mobile networks was examined by Wakikawa et al. [15]. The routing has been implemented through the ad-hoc routing algorithm or the basic NEMO networking technique. Jonsson et al. [16] developed the Mobile IP for MANET (MIPMANET) system, which uses Mobile IP to provide the Ad Hoc On-Demand Distance Vector (AODV)-based MANETs with internet connectivity between such a gateway and the MANET is a MIPMANET interworking unit.

There has been a lot of research on reasoning threats or defects in databases [10] and networking [11,12]. Only a few articles have looked at reasoning in a decentralized and automated approach for the route. SNP [13] describes why computer networks are in a specific status to their controllers. ExSPAN [14] is a database framework that enables reasoning. None of these studies addresses why those conditions or log records are missing but do not appear. Wu et al. [15]describe a weak reasoning connection strategy to solve why-not queries in SDN. The study "Why not?" [16] can monitor unfavorable reasoning links in SDN and BGP as well. Both projects should have human workers, but they must not begin immediately. SDN can receive information from a controller that collects data from sensor nodes without regard for privacy.

Broch J. et al. [17] offer a paradigm that enables a DSR-based MANET to traverse diverse connection stages with a single access point. In a MANET IP subnet, this structure only allows for one gateway. In ad hoc networks, Kock and Schmidt [18] developed emerging mobile IP routers that connect to the entire system. Tseng et al. [19] suggested a method for upgrading typical IEEE 802.11-based edge routers to include the mobility of mobile nodes, allowing the ideal of pervasive broadband wifi access to become a truth. Perkins et al. [20] utilized Mobile IP as the foundation for offering mobile users movement, then extended it to enable extra services to mobile users at the network level and higher.

In MANET, some techniques address the privacy difficulties raised by Kong et al.[14]. Anonymous On-Demand Routing (ANODR) is the opening technology to guarantee privacy in ad hoc networks during multipath routing and packet forwarding. After ANODR's work, Seys and Preneel[22] introduced the Anonymous Routing Mechanism (ARM), which employs a one-time public/private pair of keys and addresses privacy in route finding and packet forwarding. Sy et al. [23]proposed On-Demand Anonymous Router (ODAR) for reliable anonymous routing utilizing public-key cryptographic systems. Still, we require lengthy public/private key pairings to be established on every node for encrypted communication. When compared to ANODR, which may decrease routing performance, Zhang et al.[24] The Anonymous On-Demand Router (MASK) provides an AODV-like anonymously on-demand direction-finding scheme with excellent routing effectiveness.

## 3. Proposed Methodology

### 3.1. Architecture

As shown in Fig. 1, TAP3 is a protocol that consists of three phases: neighbor identification, trust estimate, and verification. It provides a safe information exchange path and limits dangerous nodes by allowing each node to authenticate and calculate trust. Nodes, cluster heads, and BS make up the network. BS provides centralized control and aids in reducing network capacity and processing requirements. For secure communication between two nodes, our proposed routing protocol employs a Modified Nearest pair-wise keys pre-distribution technique [25]. The setup server uses master keys to all servers, and for each pair of nodes (IDS, IDR), a complete set key KS, R=PRF KR(S) is created, where PRF stands for pseudo-random functional. All sensor networks in the communication limit of a new sensor node have specified keys. In addition, the Hash Message Authentication Code (HMAC) is used to ensure message integrity and validate sender authenticity [25-27].

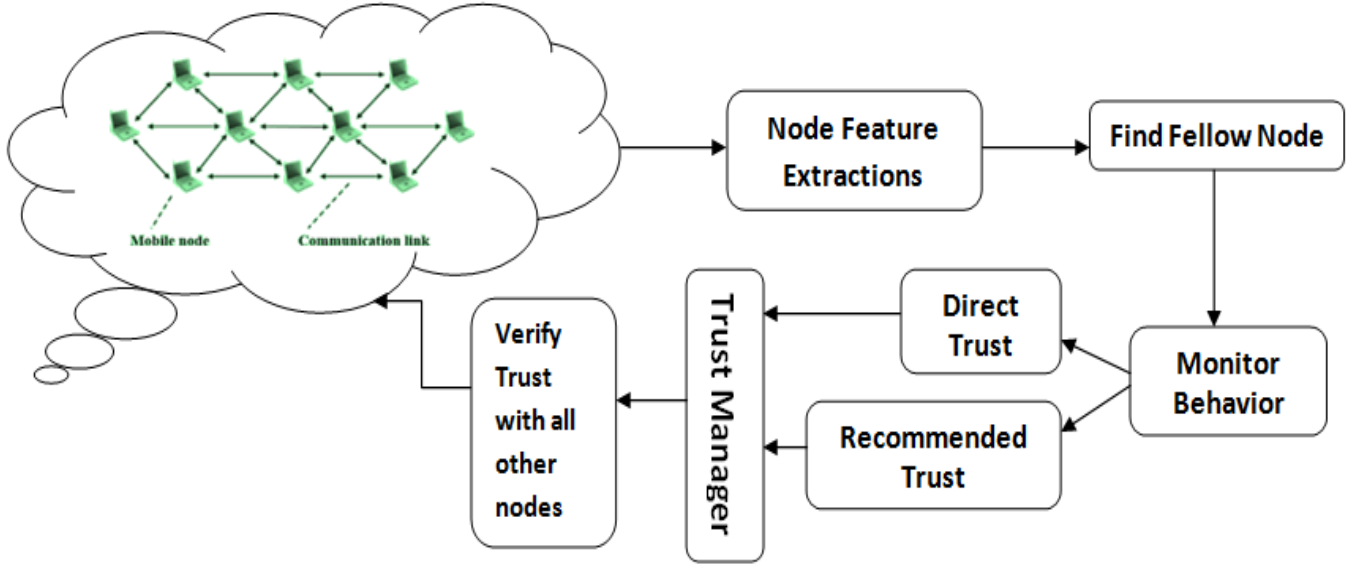

**Fig. 1 Architecture of Proposed Model**

### 3.2. Dynamic Fellow Identification

Conventional Multicast routing technology needs each control and information payload to have the target address to determine a route and recognize a fellow [28-30]. An enemy node near the start or end, or on the communications channel among the two, would be able to connect the friend nodes and maybe learn their geo-location using this basic approach. We develop a dynamic fellow identification approach based on forwarding clustering to identify a fellow without disclosing the source and destination addresses. Two fellow pseudonyms, PDi and PSi, are established in the active fellow identification scheme for the forward and backward fellows accordingly. The sending and receiving addresses of the packets are replaced by the fellow alias.

A transmitter node out an RREQ packet, including these other fake identities. When intermediate nodes get an RREQ signal, they try to decode and analyze fellow fake names to see if they are the target or "open the trapdoor," which hides the source and target addresses as described following. They create a routing information item for each fellow indicated by fellow identity PSi in RREQ because they're not the destination. When an endpoint receives an RREQ, it checks the received PDi to see if it is indeed the target. Because each node must execute a backdoor check, the efficiency of the check is critical. Using symmetric cryptography [31-33] and genuine IDs of transmitter and receiver, the first fellow fake accounts, PD1 and PS1, with forward and reverse fellows, are formed. An origin or a recipient can modify the nickname of a fellow at any moment. The forward sequence is used to produce successive fellow aliases based on the prior fellow alias to accomplish :

PS1=fnKSD (S)→PS2=fnKSD (PS1)...→PSn=fnKSD (PSn−1)

PD1=fnKSD (D)→PD2=fnKSD (PD1)...→PDn=fnKSD (PDn−1),

To the intermediary nodes, the outcome of function fn appears random. The trapdoor verification is fundamental, requiring simply the computation of a hash and a quick search for a corresponding node [34-36]. Furthermore, the trapdoor test is performed when analyzing the RREQ packet; once the fellow has been forwarded, the check is no longer necessary when forwarding the following packets. An ideal data structure, such as a binary search tree, can be utilized to increase the efficiency of the trapdoor check-in of each node.

### 3.3. Dynamic Monitoring for Malicious Identification

When senders send RREQ and target nodes react to RREP, they may all change their hash values to match the larger ones. The increasing speed of every node's OSeq and the DSeq in the RREP sent from the endpoints depends on the network's traffic status. For example, OSeq and DSeq will increase more quickly if the endpoint is dynamic on the route than if it is dormant. As a result, the DSeq threshold cannot be set to an unchanging value, and we must use dynamic, active learning to forecast the current DSeq level. The set of connections (network) condition in time slot Dti is defined by a 3D vector for Request(RREQ) and Reply(RREP) that travel across every node: yi = (yi1, yi2, yi3). Slot of time Dti is the time interval between when the source first started sending RREQ and when the final constructed route is reached. The sample estimated time is T, including N times during the route discovery process [37-40]. The time it takes for the source delivering the RREQ to obtain the RREP ACK sent from the target is used to calculate the route discovering time. yi1 is the original sequence digit (SSeq) in RREP or RREQ, yi2 is the node's unique identifier (OSeq), and yi3 is the variance between RREP and RREQ's destination address (DSeq). The mean vector is then calculated using Eq. (1):

$$\bar{y} = \frac{1}{N}\sum_{i-1}^{N} yi, \qquad (1)$$

Where N = routing discovery period

Next, The range between the input sample data and the mean vectors is calculated $\bar{y}$:365.

$$d(y) = |y - \bar{y}|^2 \qquad (2)$$

It will be evaluated as a compromised node if the range exceeds the maximum Th (which implies it may be out of bounds as a regular sequence number).

$$\begin{cases} d(y) > Th : & Malicious \\ d(y) \le Th : Normal\ Node \end{cases} \qquad (3)$$

From the training data set, the range with the highest value is retrieved as Th:

$$Th = d(y1),\ \ I = \max(d(yi)) \qquad (4)$$

'I' is the i value that optimizes the functionality d. (xi). To observe if the sequence digit vector is appropriate, we can use Eq. (3) to calculate the sequence digit threshold (4). We presume no malicious node exists at the start of the learning time interval. We let the nodes practice training from the first period T1 and then utilize the outcome for the sequence number judgment for the next time interval T2. If T2 is considered at steady levels, it will add the new data set to the training sample, and the old training dataset will be updated.

### 3.4. TAP3 Algorithm

S and D stand for transmitter and recipient, respectively, and Mi stands for intermediate routers (M1; M2; ...Mn). We offer three approaches to provide a reliable identification process: target log checking, active threat detection, and passive threat detection [41-44]. Which method we should employ in the next stage is determined by the result of the target log verification method. In Algorithm1's target log checking, we first determine the constraints that the origin is inferring for the target. The necessary set of rule ri is ri(left), while the resultant part of rule ri is ri(right). Then we utilize MHT to double-check the destination's log. We will use Algorithm2 to detect active attacks if the confirmation outcome does not match. Meanwhile, assuming the result is correct, we proceed to Algorithm3 to detect passive attacks.

**Algorithm 1 Fellow Node Identification**
Input: $\tau c$, record entries; $rc{\rightarrow}Dt$, the policy for determining the endpoint from the origin
Output: *outcome*, indicating whether or not the log is lying on the target is right *(outcome* = FELLOW) or not *(outcome* ≠ FELLOW).
Step 1: for $ri\ \varepsilon$ rc→ dt, do
Step 2:    if $ri(lhs) \subseteq \tau_C$ then
Step 3: $map_i$.put *(n(lhs), n(rhs));*
Step 4:    finish if
Step 5: finish for
Step 6: $Collection_i\_Record{\leftarrow}map.value();$
Step 7: for $i \leftarrow 0$ to $Collection\_Record.size()$ do
Step 8:    *outcome* ←Hash_Verify*(Dt, Collection_i Record);*
Step 9:    if *outcome* ≠ FELLOW then
Step 10: break;
Step 11:    finish if
Step 12: finish for
Step 13: return *outcome*

When the outcome in Algorithm 2 is not matching, it signifies that the channel link is under various attacks [45-47]. All intermediary gateways are stored in the intermediated route list variable (as the same case in Algorithm3). We initially filter out the rules that the origin elucidates the intermediary endpoints in active threat detection. The bit used to indicate that the intermediary routers' verification was complete is the flag bit. We know the node completed all checks if Flag = 0. To identify the intruder, we reverse-engineer the intermediary routers. We may tell that the endpoint has declared the record false if mn is authentic. When Algorithm 2 is complete, we can produce the active assailant endpoint.

**Algorithm 2 Dynamic Attack Detection**

Input: *Intermediary_route_list* (n1, n2, n3...n$_n$), containing intermediary routers; $\tau c$, log entries; rc→M, the policy that start node identifies the intermediary nodes.

Outcome: The Fake node initiated the dynamic assault.

Step1: for *ri ε* rc→M, do
Step2:    if *ri(lhs)* ⊆ $\tau_{c,}$ then
Step3:   map$_i$.put *(ri(lhs), ri(rhs))*
Step4:     finish if
Step5: finish for
Step6: *Collection_Record* ←*map.values();*
Step7: for m ← n to 1
Step8:   flag_bit ← 0;
Step9:    for *i* ← 0 to *Collection_Record.Size()* do
Step10: outcome ← Hash_Verify(m$_j$, *Collection$_i$_Record);*
Step11: if outcome ≠ FELLOW then
Step12: flag_bit ←1;
Step13: break;
Step14: finish if
Step15: finish for
Step16: if flag_bit=0 then
Step17: if *m = n,* then
Step18: return *Target;*
Step19: else
Step20: return *Nm*
Step21: finish if
Step22: finish if
Step23: finish for

As soon as the outcome of Algorithm1 is the same, we recognize we must investigate whether the linkage was subjected to a passive assault [49-53]. We, moreover, learned the source's rules to determine the intermediate nodes from the target. Then we go through all intermediate nodes and double-check the predicted logs. Finally, when detecting them, we add faulty nodes to the Malicious_list(). As it will examine more records and visit all intermediary nodes, Algorithm3 may take longer than Algorithm2.

**Algorithm 3 Submissive Attack Detection**

Input: *intermediary_route_list* (n1, n2, n3... n$_n$), containing intermediary routers; $\tau c$, log entries; $\tau d$, target node log entries rc+D→M, the policy that start node identifies the intermediary nodes.

Outcome: *Fakenode_list(),* the Fake nodes list initiating the submissive attack.

Step1: for *ri ε* rc +D→M, do
Step2:   if *(ri(lhs)* ⊆ $\tau_c \cup \tau_D$), then
Step3: *map$_i$.put (ri(lhs), ri(rhs));*
Step4: finish if
Step5: finish for
Step6: *Collection_Record* ← *map.values();*
Step7: for *j* ← 1 to N do

Step8:    for i ← 0 to *Collection_Record.Size()* do
Step9: *outcome* ← MHT_Verify(nj, *Collection$_i$_Record);*
Step10: if outcome ≠ FELLOW then
Step11: *Fakenode_list.* add(Nj);
Step12: break;
Step13: finish if
Step14: finish for
Step15: finish for
Step16: return *Fakenode_list*

## 4. Experimental Results

This part looks at how TAP3 affects navigation and data transmission rate. The ns2 simulator [9] was used to run our simulation. We examine the impact of TAP3 in situations when numerous pathways are recognized, and each packet on a stream may travel in a diverse direction. We adopt ad hoc on-demand multipath distance vector routing (AOMDV) as a basic multipath routing [17]. To use TAP3, we updated it to develop S-MPRF, a version that employs a stable station pseudonym. The simulation environment is summarized in the following table.

**Table 1. Simulation Parameters**

| Simulation Period | 1000 sec |
|---|---|
| No of nodes | 100 |
| Simulation Area | 800X800 |
| Simulation Speed | 25 m/sec |
| Type of Model | Random Waypoint Model |
| Packet length | 256 bytes |
| Traffic pattern | 12 CBR/UDP connections (4 packets/s) |

We evaluated packet delivery ratio (PDR), point-to-point transmission interruption (delay), and routing overhead with various timeslots and a random waypoint model. With changing demographics, various multi-path forwarding degrades the PDR.

**Table 2. PDR Rate between TAP3 and MTRF and S-MTRF**

| S.No | Pause Time (s) | PDR (%) | | |
|---|---|---|---|---|
| | | **MTRF** | **S-MTRF** | **TAP3** |
| 1 | 0 | 90.5 | 93 | 95 |
| 2 | 50 | 88 | 91 | 93 |
| 3 | 100 | 90 | 92 | 95.5 |
| 4 | 150 | 90 | 92.5 | 95 |
| 5 | 200 | 89.5 | 91.5 | 93.5 |
| 6 | 250 | 90.5 | 92.5 | 94 |
| 7 | 300 | 92 | 94 | 96 |

As a result, packets are more susceptible to connection breakdown or network jamming because every packet follows a different route.

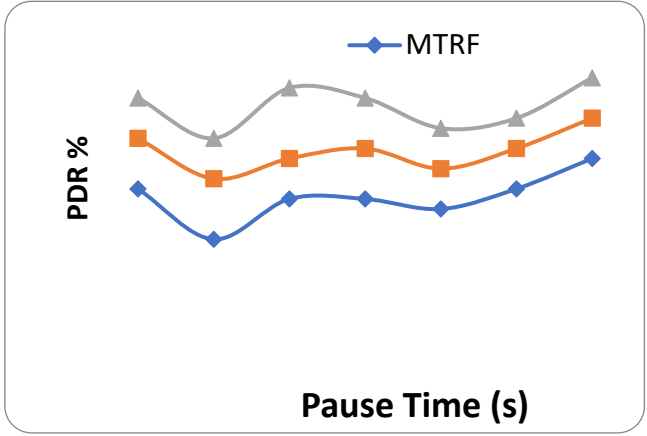

**Fig. 2a Packer Delivery Ratio (%)**

Table. 2 and Fig. 2 (a) demonstrate that the PDR is high for TAP3 and is reduced by 3% and 5% in various multi-route routing protocols, such as S-MPRF and MPRF, correspondingly. This outcome indicates that modifying node pseudonyms has a minor influence. The fact that each stream has multiple paths that can break tends to increase the forwarding overhead considered necessary to resolve these failures. MPRF has a 42 percent higher routing overhead than TAP3, as seen in Fig. 2 (c) and Table. 4.

**Table 3. Average Packet Delay (s)**

| S.No | Break Time (s) | PDR (%) | | |
|------|------|------|------|------|
| | | **MTRF** | **S-MTRF** | **TAP3** |
| 1 | 0 | 0.14 | 0.15 | 0.11 |
| 2 | 50 | 0.15 | 0.13 | 0.12 |
| 3 | 100 | 0.13 | 0.15 | 0.1 |
| 4 | 150 | 0.15 | 0.13 | 0.1 |
| 5 | 200 | 0.13 | 0.12 | 0.09 |
| 6 | 250 | 0.12 | 0.13 | 0.11 |
| 7 | 300 | 0.12 | 0.125 | 0.085 |

Data packets deliver on the shortest route in conventional routing methods. Data packets are dynamically spreading across various pathways with TAP3. The end-to-end packet latency will rise because specific pathways will be longer than the shortest. Compared to TAP3, Tab. 3 and Fig. 2 (b) reveals a 50% rise in packet deliverance time compared to MPRF and S-MPRF.

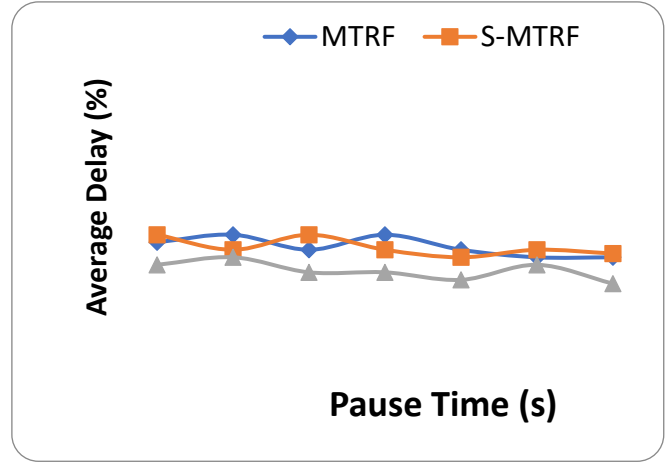

**Fig. 2b Average Packet Delay (%)**

**Table 4. Average Packet Delay (s)**

| S.No | Break Time (s) | PDR (%) | | |
|------|------|------|------|------|
| | | **MTRF** | **S-MTRF** | **TAP3** |
| 1 | 0 | 2.4 | 2.3 | 1.74 |
| 2 | 50 | 2.8 | 2.75 | 2.2 |
| 3 | 100 | 2.5 | 2.6 | 1.72 |
| 4 | 150 | 2.3 | 2.35 | 1.5 |
| 5 | 200 | 2.35 | 2.4 | 1.52 |
| 6 | 250 | 2.4 | 2.3 | 1.55 |
| 7 | 300 | 2.45 | 2.32 | 1.5 |

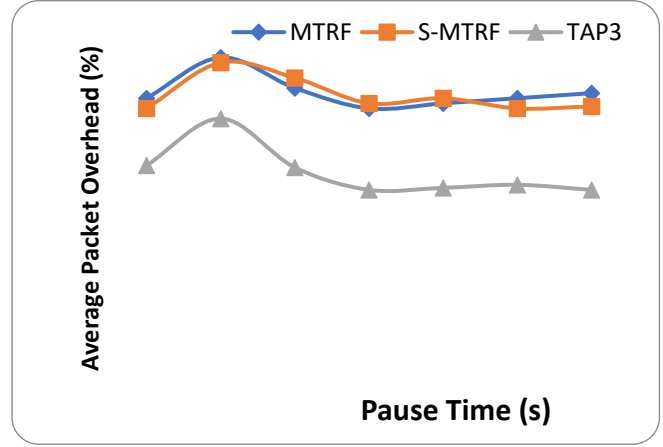

**Fig. 2c Average Packet Overhead (%)**

## 5. Conclusion

When used with a base station to connect to the internet, the TAP3 addresses privacy and security concerns in wireless ad hoc networks. The design adapts components such as monitoring, trustworthiness estimation, and reputation to supervise and decide the trust and reputation of network nodes. An observer determines the node state based on these variables and notifies other participants in the system if the node is malfunctioning (outlier or malicious). The TAP3 combines the fellow identification and trust estimation system with privacy-preserving routing to meet privacy and security requirements. As a result, the suggested scheme has significance in the integrated context of wireless ad hoc networks. TAP3's performance is demonstrated via simulated results. For a limited increase of active nodes, the suggested scheme is appropriate. However, when the number of active nodes in a network surpasses a certain threshold, the mobile node energy usage is disproportionately large. As a result, the scalability element must be considered alongside existing principles in the future. In terms of latency, delivery ratio, and overhead rate, TAP3 outperforms MTRF and S-MTRF protocols.

## References

[1]     D Axiotis, T Al-Gizawi, K Peppas, E Protonotarios, F Lazarakis, C Papadias, P Philippopoulos, "Services in Interworking 3G and WLAN Environments," *IEEE Wireless Communication*, vol. 11, no. 5, pp.14–20, 2004.

[2]     M Lott, M Siebert, S Bonjour, D Von Hugo, M Weckerle, "Interworking of WLAN and 3G Systems," *IEEE Proceeding Communication*, vol. 151, no. 5, pp.507–513, 2004.

[3]     B. Murugeshwari, D. Selvaraj, K. Sudharson and S. Radhika, "Data Mining With Privacy Protection Using Precise Elliptical Curve Cryptography," *Intelligent Automation & Soft Computing*, vol. 35, no . 1 , pp.839–851, 2023.

[4]     K. Sudharson, and V. Parthipan, "A Survey on ATTACK – Anti Terrorism Technique for Adhoc Using Clustering and Knowledge Extraction," *Advances in Computer Science and Information Technology*. Computer Science and Engineering. CCSIT 2012. Lecture Notes of the Institute for Computer Sciences, Social Informatics and Telecommunications Engineering, Springer, Berlin, Heidelberg, vol. 85, pp. 508-514, 2012.

[5]     N Komninos, DD Vergados, C Douligeris, "A Two-Step Authentication Framework for Mobile Ad Hoc Networks," *China Commun J*, vol. 4, no. 1, pp.28–39, 2007.

[6]     D. Dhinakaran, P.M. Joe Prathap, D. Selvaraj, D. Arul Kumar, and B. Murugeshwari, "Mining Privacy-Preserving Association Rules Based on Parallel Processing in Cloud Computing," *International Journal of Engineering Trends and Technology*, vol. 70, no. 30, pp.284-294, 2022. doi: 10.14445/22315381/IJETT-V70I3P232.

[7]     K El Defrawy, G Tsudik, "Privacy-Preserving Location-Based on-Demand Routing in Manets," *IEEE Journal on Selected Areas Communications*, vol. 29, no. 10, pp.1926–1934, 2011.

[8]     A Loay, K Ashfaq, G Mohsen, "A Survey of Secure Mobile Ad Hoc Routing Protocols," *IEEE Communications Survey and Tutorials*, vol. 10, no. 4, pp.78–93, 2008.

[9]     A Irshad, M Shafiq, A Rahman, S Khurram, M Usman, E Irshad, "A Secure Interaction Among Nodes From Different MANET Groups Using 4G Technologies," *In International Conference on Emerging Technologies*, Islamabad, pp.476–481, 2009

[10]    D Shuo, "A Survey on Integrating Manets With the Internet: Challenges and Designs," *Computer Communication*, vol. 31, no. 14, pp.3537–3551, 2008.

[11]    C Imrich, C Marco, L Jennifer, "Mobile Ad Hoc Networking: Imperatives and Challenges," *Ad Hoc Network*, vol. 1, no. 1, pp.13–64, 2003.

[12]    Y Sun, E Royer, CE Perkins, "Internet Connectivity for Ad Hoc Mobile Networks," *International Journal of Wireless Information Network*, vol. 9, no. 2, pp.75–78, 2002.

[13]    I Teranishi, J Furukawa, K Sako, "K-Times Anonymous Authentication," *in Proceedings of ASIACRYPT* (Jeju Island, 2004) pp.308–322, 2004.

[14]    J Kong, X Hong, "ANODR: Anonymous on-Demand Routing With Untraceable Routes for Mobile Ad-Hoc Networks," *In Proceedings of 4th ACM International Symposium on Mobile Ad Hoc Networking and Computing* (Annapolis, 2003) vol. 291–302, 2003.

[15]    R Wakikawa, H Matsutani, R Koodli, A Nilsson, J Murai, "Mobile Gateways for Mobile Ad-Hoc Networks with Network Mobility Support*," In Proceedings of 4th International Conference on Networking* (Reunion Island, France, 2005) pp.17–21, 2005.

[16]    U Jonsson, F Alriksson, T Larsson, P Johansson, J Maguire, "MIPMANET Mobile IP for Mobile Ad Hoc Networks (Paper Presented in the 1st Annual Workshop on Mobile and Ad Hoc Networking and Computing, Boston, MA, pp.75–85, 2000.

[17]    J Broch, DA Maltz, DB Johnson, "Supporting Hierarchy and Heterogeneous Interfaces in Multi-Hop Wireless Ad Hoc Networks," *in Proceedings of the International Symposium on Parallel Architectures, Algorithms and Networks* , pp.370–375, 1999.

[18]    BA Kock, JR Schmidt, "Dynamic Mobile IP Routers in Ad Hoc Networks, *(Paper Presented in the International Workshop on Wireless Ad-Hoc Networks, Netherlands*, pp.130–134, 2004.

[19]    Y Tseng, C Shen, W Chen, "Integrating Mobile IP With Ad Hoc Networks," *IEEE Computer Society*, vol. 36, no. 5, pp.48–55, 2003.

[20]    J CE Perkins, NY Thomas, Yorktown Heights, Mobile-IP, "Ad-Hoc Networking, and Nomadicity," *In Proceedings of 20th International Conference on Computer Software and Applications Conference*, pp.472–476, 1996.

[21] D. Dhinakaran, D. A. Kumar, S. Dinesh, D. Selvaraj, and K. Srikanth, "Recommendation System for Research Studies Based on GCR," *2022 International Mobile and Embedded Technology Conference (MECON),* Noida, India, pp.61-65, 2022. Doi: 10.1109/MECON53876.2022.9751920.

[22] MG Reed, PF Syverson, DM Goldschlag, "Anonymous Connections and Onion Routing," *IEEE J Sel Area Comm*, vol. 16, no. 4, pp.482–494, 1998.

[23] S Seys, B Preneel, "ARM: Anonymous Routing Protocol for Mobile Ad Hoc Networks," In Proceedings of the International Conference on Advanced Information Networking and Applications , *IEEE Computer Society*, Washington, DC, Switzerland, pp.145–155, 2009.

[24] D Sy, R Chen, L Bao, "ODAR: on-Demand Anonymous Routing in Ad Hoc Networks," *In Proceedings of the 3rd IEEE International Conference on Mobile Ad-Hoc and Sensor Systems* (Vancouver, BC, 2006) , pp.267–275, 2006.

[25] Y Zhang, W Liu, W Lou, Y Fang, "MASK: Anonymous on-Demand Routing in Mobile Ad Hoc Networks," *IEEE Trans Wireless Comm*,  vol. 5, no. 9, pp.2376–2385, 2006.

[26] Balasubramaniam, Murugeshwari & Daniel, Raphael & Raghavan, Singaravelu, "Metamaterial Inspired Structure with Offset-Fed Microstrip Line for Multi Band Operations," *Progress in Electromagnetics Research* M, vol. 82, pp.95-105, 2019.

[27] D. Dhinakaran  and P.M Joe Prathap, "Protection of Data Privacy From Vulnerability Using Two-Fish Technique With Apriori Algorithm in Data Mining," *Journal of Supercomputing*, 2022. https://doi.org/10.1007/S11227-022-04652-8.

[28] S. Arun, and K. Sudharson, "DEFECT: Discover and Eradicate Fool Around Node in Emergency Network Using Combinatorial Techniques," *Journal of Ambient Intelligence and Humanized Computing*, pp.1-12, 2020. Doi: Https://Doi.Org/10.1007/S12652-020-02606-7.

[29] J. Aruna Jasmine, V. Nisha Jenipher, J. S. Richard Jimreeves, K. Ravindran, and D. Dhinakaran, "A Traceability Set Up Using Digitalization of Data and Accessibility," *2020 3rd International Conference on Intelligent Sustainable Systems (ICISS),* pp.907-910, 2020.

[30] H Choi, W Enck, J Shin, P Mcdaniel, T La Porta, "ASR: Anonymous and Secure Reporting of Traffic Forwarding Activity in Mobile Ad Hoc Networks," *Springer Link. Wireless Netw.* vol. 15, no. 4, pp.525–539, 2009.

[31] B Zhu, K Ren, L Wang, "Anonymous Misbehavior Detection in Mobile Ad Hoc Networks,"  *in Proceedings of 28th International Conference on Distributed Computing Systems Workshops* (IEEE Computer Society, Beijing,), pp.358–363, 2008.

[32] B. Murugeshwari, K. Sarukesi and C. Jayakumar, "An Efficient Method for Knowledge Hiding Through Database Extension," *Test Conference, International*, vol. 2010, pp. 342-344. 10.1109/ITC.2010.93.

[33] K. Sudharson, and V. Parthipan, "SOPE: Self-Organized Protocol for Evaluating Trust in MANET Using Eigen Trust Algorithm," *2011 3rd International Conference on Electronics Computer Technology*, Kanyakumari, India, pp.155-159, 2011.

[34] B. Murugeshwari, R.S. Daniel and S. Raghavan, "A Compact Dual Band Antenna Based on Metamaterial-Inspired Split Ring Structure and Hexagonal Complementary Split-Ring Resonator for ISM/Wimax/WLAN Applications," *Appl. Phys. A* , vol. 125, no. 628 , 2019.Https://Doi.Org/10.1007/S00339-019-2925-X.

[35] D. Dhinakaran, and P.M. Joe Prathap, "Ensuring Privacy of Data and Mined Results of Data Possessor in Collaborative ARM, Pervasive Computing and Social Networking," *Lecture Notes in Networks and Systems*, Springer, Singapore, vol. 317 , pp.431 – 444, 2022. DOI: 10.1007/978-981-16-5640-8_34.

[36] J Pan, J Li, "MASR: An Efficient Strong Anonymous Routing Protocol for Mobile Ad Hoc Networks," *in Proceedings of the International Conference on Management and Service Science* (Wuhan), pp.1–6, 2009.

[37] M Gunasekaran, K Premalatha, "TEAP: Trusted-Enhanced Anonymous on Demand Routing Protocol for Mobile Ad Hoc Networks," *IET Inf Secur*, vol. 7, no. 3, pp.203–211, 2012.

[38] K. Sudharson and S. Arun, "Security Protocol Function Using Quantum Elliptic Curve Cryptography Algorithm," *Intelligent Automation & Soft Computing*, vol. 34, no. 3, pp.1769–1784, 2022.

[39] D Johnson, A Menezes, S Vanstone, "The Elliptic Curve Digital Signature Algorithm (ECDSA)," *International Journal of Information Security*, vol. 1, no. 1, pp.36–63, 2001.

[40] Q Huang, D Jao, HJ Wang, "Applications of Secure Electronic Voting to Automated Privacy Preserving Troubleshooting (ACM, New York)," pp.68–80, 2005.

[41] K. Sudharson, M. Akshaya, M. Lokeswari and K. Gopika, "Secure Authentication Scheme Using CEEK Technique for Trusted Environment," *2022 International Mobile and Embedded Technology Conference (MECON)*, Noida, India, pp.66-71, 2022.

[42] Murugeshwari, "Preservation of the Privacy for Multiple Custodian Systems With Rule Sharing," *Journal of Computer Science*, vol. 9 pp.1086-1091, 2013. 10.3844/Jcssp.2013.1086.1091.

[43] Li, T., Ma, J., & Sun, C., "Netpro: Detecting Attacks in Manet Routing With Provenance and Verification," *Science China Information Sciences,* vol. 60, no. 11, 2017.

[44] Rathee, G.; Saini, H, "Secure Handoff Technique With Reduced Authentication Delay in Wireless Mesh Network," *International Journal Advanced Intelligence Paradigms*, vol. 13 , pp.130–154, 2019.

[45] D. Dhinakaran, and P. M. Joe Prathap, "Preserving Data Confidentiality in Association Rule Mining Using Data Share Allocator Algorithm," *Intelligent Automation & Soft Computing*, vol. 33, no. 3, pp.1877–1892, 2022. DOI:10.32604/Iasc.2022.024509.

[46] N. Partheeban, K. Sudharson, and P.J. Sathish Kumar, "SPEC- Serial Property Based Encryption for Cloud," *International Journal of Pharmacy & Technology*, vol. 8, no. 4, pp.23702-2371, 2016.

[47] Jhaveri, R.H.; Ramani, S.V.; Srivastava, G.; Gadekallu, T.R.; Aggarwal, V, "Fault-Resilience for Bandwidth Management in Industrial Software-Defined Networks," *IEEE Transactions on Network. Science and Engineering*, vol. 8, pp.3129–3139, 2021.

[48] K. Sudharson, A. M. Sermakani, V. Parthipan, D. Dhinakaran, G. Eswari Petchiammal and N. S. Usha, "Hybrid Deep Learning Neural System for Brain Tumor Detection," *2022 2nd International Conference on Intelligent Technologies (CONIT)*, pp.1-6, 2022. Doi: 10.1109/CONIT55038.2022.9847708.

[49] P Aishwarya Naidu, Satvik Vats,Pooja Chadha, Rajeshwari K, "Vehicular Ad-Hoc Networks and Associated Risks, " *International Journal of P2P Network Trends and Technology*, vol. 10, no. 3, pp.18-25, 2020.

[50] Maddikunta, P.K.R.; Srivastava, G.; Gadekallu, T.R.; Deepa, N.; Boopathy, P, "Predictive Model for Battery Life in Iot Networks," *IET Intelligent Transport Systems*, vol. 14, pp.1388–1395, 2020.

[51] Wu, T.Y.; Lee, Z.; Yang, L.; Luo, J.N.; Tso, R, "Provably Secure Authentication Key Exchange Scheme Using Fog Nodes in Vehicular Ad Hoc Networks," *Journal of Supercomputing,* vol. 77, pp.6992–7020, 2021.

[52] P Jianli, P Subharthi, J Raj, "A Survey of the Research on Future Internet Architectures," *IEEE Commun Mag*, vol. 49, no. 7, pp.26–36, 2011.

[53] FM Abduljalil, SK Bodhe, "A Survey of Integrating IP Mobility Protocols and Mobile Ad Hoc Networks," *IEEE Communication Survey Tutorials*, vol. 9, no. 1, pp.14–30, 2007.

[54] Jeyalakshmi, C. & Balasubramaniam, Murugeshwari & Karthick, M, "HMM and K-NN Based Automatic Musical Instrument Recognition," pp.350-355, 2018. 10.1109/I-SMAC.2018.8653725.